\documentstyle{elsart}

\input psfig

\begin{document}

\begin{frontmatter}
\title{Has the GZK suppression been discovered?}

\author[princeton]{John N. Bahcall\thanksref{jnbmail}} and \author[weizmann]{Eli Waxman\thanksref{ewmail}}
\address[princeton]{School of Natural Sciences, Institute for Advanced Study, Princeton, NJ 08540, USA}
\address[weizmann]{Physics Faculty, Weizmann Institute of Science, Rehovot
76100, Israel}
\thanks[jnbmail]{E-mail: jnb@sns.ias.edu}
\thanks[ewmail]{E-mail: waxman@wicc.weizmann.ac.il}

\begin{abstract}

The energy spectra of ultra high energy cosmic rays reported by the AGASA, Fly's Eye, Haverah Park,
HiRes, and Yakutsk experiments are all shown to be in agreement with each other for energies below
$10^{20}$~eV (after small adjustments, within the known uncertainties, of the absolute energy scales). The
data from HiRes, Fly's Eye, and Yakutsk are consistent with the expected flux suppression above $5\times
10^{19}$~eV due to interactions of cosmic rays with the cosmic microwave background, the
Greisen-Zatsepin-Kuzmin (GZK) suppression, and are inconsistent with a smooth extrapolation of the
observed cosmic ray energy spectrum to energies $> 5\times 10^{19}$ eV. AGASA data show an excess of
events above $10^{20}$~eV, compared to the predicted GZK suppression and to the flux measured by the
other experiments.
\end{abstract}
\end{frontmatter}

\begin{figure}[!tb]
\centerline{(a)\psfig{figure=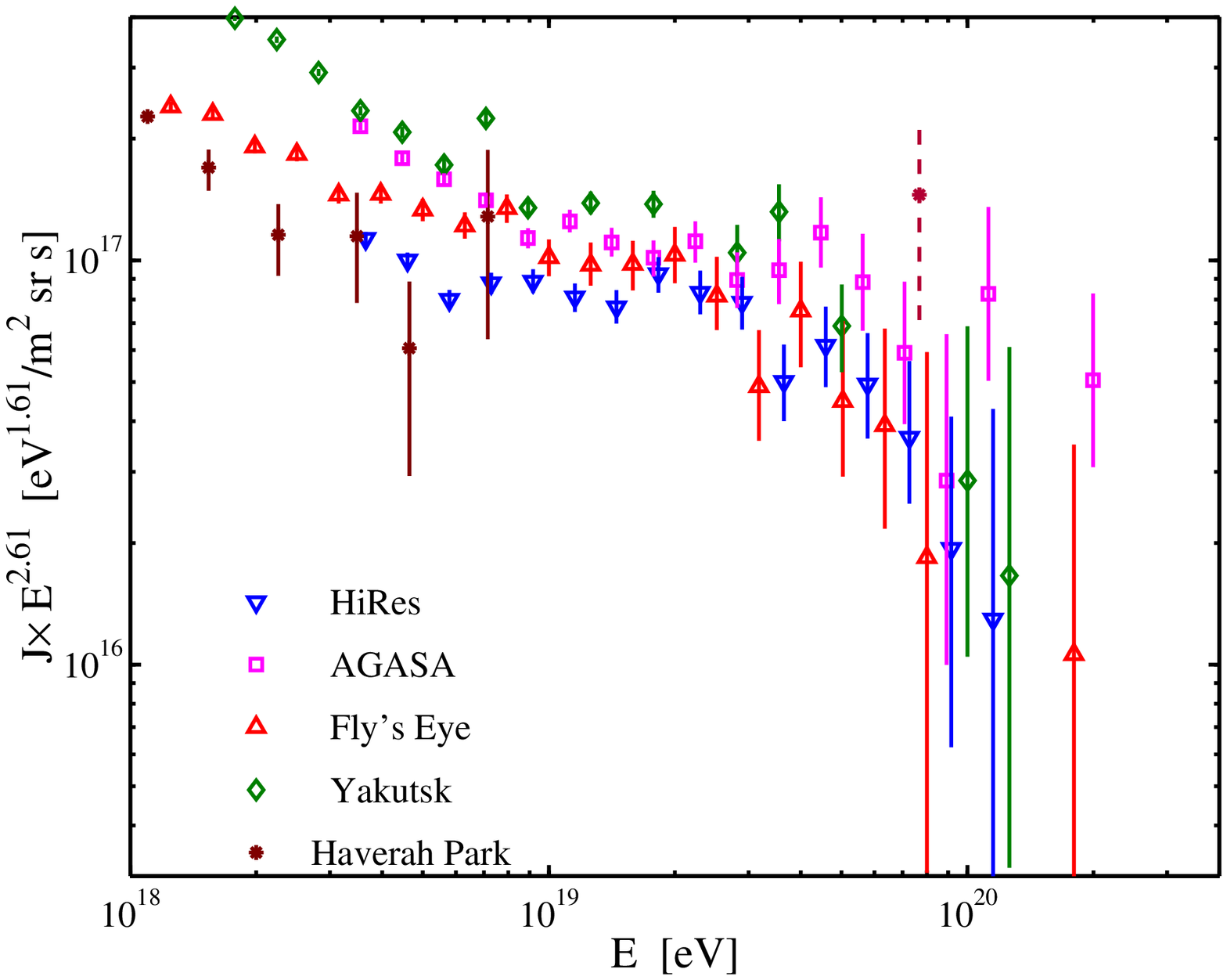,width=3.3in}}
\centerline{(b)\psfig{figure=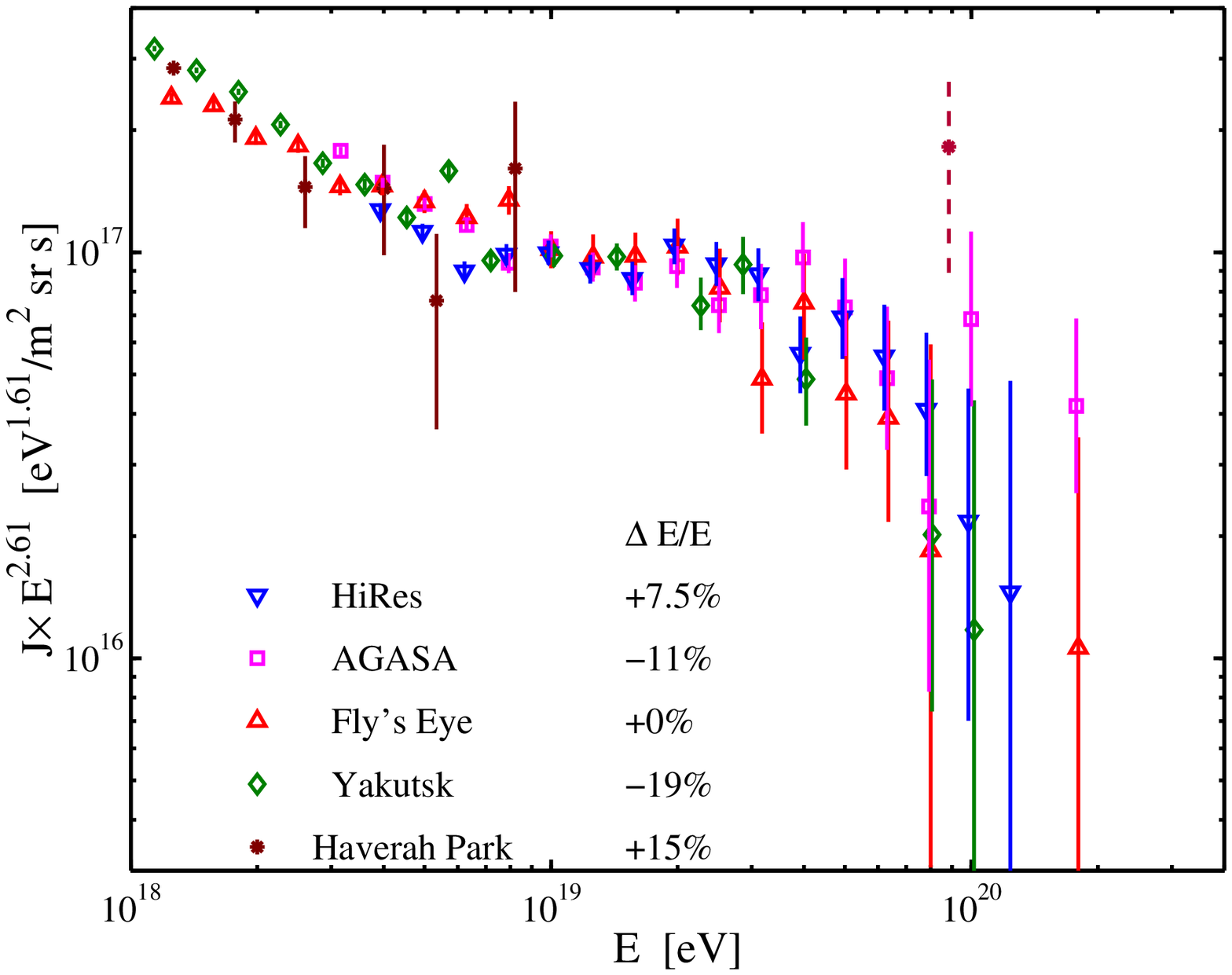,width=3.3in}} \caption{Currently available data on
the highest energy cosmic rays. The quantity $J$ is the differential energy flux, $d\phi/dE$, per unit
time per unit area per steradian. Panel (a) shows the data as published by the five experimental
collaborations: AGASA \cite{agasa}, Fly's Eye \cite{fly}, Haverah Park \cite{HP}, HiRes \cite{HiRes}, and
Yakutsk \cite{yakutsk}. Panel~(b) shows the data after adjusting the absolute energy calibrations of the
various experiments so as to bring the results from the different experiments into agreement at $10^{19}$
eV. For specificity, the Fly's Eye absolute energy scale was adopted as the standard. The fractional
shifts in absolute energy scale, $\Delta E/E$, shown in the figure, are all well within the published
systematic errors in the energy scale. \label{fig:beforeafter}}
\end{figure}

\section{Introduction}
\label{sec:introduction}

We analyze the observed spectrum of ultra-high energy cosmic rays. We find two main results: (i) The
energy spectra reported by the AGASA, Fly's Eye, Haverah Park, HiRes and Yakutsk experiments are all in
good agreement for energies below $10^{20}$ eV, and (ii) All the data are consistent with a GZK
suppression except for the AGASA points above $10^{20}$ eV. Our principal conclusion from these two
results is that standard physics, including the GZK suppression, is sufficient to explain all of the
existing data on UHE cosmic rays.

For any theoretical model in which the GZK suppression is present, the assumed intrinsic spectrum
produced by the UHE cosmic-ray sources influences the energy spectrum predicted by the model. Our
conclusion that the data are consistent with a GZK suppression implies that the observed spectrum is
consistent with model predictions for a plausible intrinsic energy spectrum. In particular, we show that
the observed spectrum is consistent with that expected for a GZK suppression of the flux produced by a
simple cosmological distribution of sources, each source producing high energy protons with a spectrum
$dN/dE_p\propto E_p^{-2}$ characteristic for collisionless shock acceleration.

Before entering into any details, we will summarize and compare in
this introduction the data that are available from different
collaborations that measure the spectrum of ultra high energy
cosmic rays.

\subsection{Summary of available data}
\label{subsec:availabledata}

Figure~\ref{fig:beforeafter} is a "Before-After" figure of the
currently available data on the highest energy cosmic rays
(energies $> 10^{18}$ eV). In Figure~\ref{fig:beforeafter}a (the
"Before" version of the figure), the data are plotted, together
with their flux error bars, as they have been published by the
five experimental collaborations: AGASA~\cite{agasa}, Fly's
Eye~\cite{fly},  Haverah Park \cite{HP}, HiRes~\cite{HiRes}, and
Yakutsk \cite{yakutsk}. The Haverah Park data have recently been
re-analyzed using modern numerical simulations of air-shower
development~\cite{HP}. The reanalysis resulted in significant
changes of inferred cosmic-ray energies compared to previously
published results (\cite{Watson91} and references quoted therein).
The data points for the  Haverah Park measurements that are shown
in Fig.~\ref{fig:beforeafter} are based on this improved analysis,
which is available only at energies $<10^{19}$~eV\footnote{A
single flux point at $\sim 7\times10^{19}$~eV is shown in
Figure~\ref{fig:beforeafter}a, but this point is based on a
preliminary analysis of 4 events that are chosen by different cuts
than those applied for the lower energy data. The energy
uncertainty for the point at $\sim 7\times10^{19}$~eV is
significantly larger than the estimated uncertainties for lower
energy points~\cite{HP}. Therefore, the point at  $\sim
7\times10^{19}$~eV is shown in Figure~\ref{fig:beforeafter}a only
for completeness; it is not used elsewhere in our analysis because
the Haverah Park collaboration has described this point as
preliminary.}.

The most striking feature of Figure~\ref{fig:beforeafter}a is that
the experimental results differ greatly among themselves (by
factors $\sim2$) even in the region $10^{18}$~eV $< E < $
$2\times10^{19}$~eV, where the quoted error bars from each
experiment are very small. In addition, the higher AGASA flux
reported above $2\times10^{20}$~eV stands out above the scatter in
the different experimental measurements.

Figure~\ref{fig:beforeafter}b (the "After" version of our "Before-After" figure) shows a dramatically
different representation of the available data. With small adjustments in the absolute energy scales, all
of the measured fluxes are seen to be in agreement at energies below $10^{20}$ eV. In constructing
Figure~\ref{fig:beforeafter}b, we have adjusted the absolute energy calibrations within the error bars
published by the experimental collaborations. We chose the shifts so as to bring the different measured
fluxes into agreement at $10^{19}$~eV. The energy shifts can be accomplished in five equivalent ways,
depending upon which one of the five energy scales is unaltered. For Figure~\ref{fig:beforeafter}b, the
Fly's Eye energy scale was unaltered and  we adjusted the AGASA energy scale by -11\%, Haverah Park by
+15\%, HiRes by +7.5\%, and Yakutsk by -19\%. All shifts are well within the published systematic errors.

Figure~\ref{fig:beforeafter} illustrates visually our two main points. First, all of the currently
available data on high energy cosmic rays are in agreement within their quoted errors for energies
between $2\times 10^{18}$~eV and $10^{20}$~eV. Second, three of the four data sets available above
$10^{19}$~eV, HiRes, Fly's Eye, and Yakutsk, all show evidence for a turnover of the energy spectrum for
energies above $5\times10^{19}$~eV. This turnover, we shall show later, is highly significant
statistically and is consistent with what one would expect from a simple model that includes the GZK
effect.  Above $10^{20}$~eV,  the reported AGASA fluxes are higher than the fluxes measured in other
experiments. It is these high AGASA fluxes alone that have led to the widespread impression that
measurements of ultra-high energy (UHE) cosmic rays (energies $>10^{19}$~eV) do not show evidence for a
GZK effect.

\subsection{What does it all mean?}
\label{subsec:whatmean}

What can one make of the results shown in the Before-After Figure~\ref{fig:beforeafter}?  There are two
simple possibilities. First, the excellent agreement shown in Figure~\ref{fig:beforeafter}b among the
different experiments could be accidental. According to this interpretation, the small adjustments made
in the energy scales are not physically motivated and the real situation is somehow much more
complicated. It is just a fluke that all of the adjusted energy spectra line up together so well below
$10^{20}$~eV. This interpretation is certainly possible. In the present paper, however, we shall choose a
different interpretation of Figure~\ref{fig:beforeafter}b. We shall suppose that the excellent agreement
of the adjusted energy spectra reveals a good approximation to the true shape of the UHE cosmic ray
energy spectra. We shall now explore the consequences of this assumption.

We stress that the distinction between the two possibilities for
interpreting Figure~\ref{fig:beforeafter}b can only be settled by
a new generation of precise and high statistics measurements of
the UHE cosmic ray spectrum. Fortunately, the Auger experiment,
currently under construction~\cite{auger}, is expected to provide
the necessary precision and statistics. The Telescope Array
experiment~\cite{TA}, currently under planning, may also provide
similar precision and statistics.

We first describe in Section~\ref{sec:model} the model we use and then in Section~\ref{sec:comparison} we
compare the model predictions with observations of the UHE cosmic ray energy spectrum. We summarize our
main conclusions in Section~\ref{sec:discussion}.

\section{A simple two-component model}
\label{sec:model}

We describe in this section a simple two-component model for the
energy spectrum of the highest energy cosmic rays. The Galactic
component is taken from observations of the Fly's Eye group. The
two input parameters for the extra-galactic component (the rate of
energy deposition in cosmic rays and the shape of the initial
spectrum) were originally suggested by the
idea~\cite{W95,Vietri95,MU95} that gamma-ray bursts (GRBs) are the
source of UHE cosmic rays. However, any postulated cosmologically
distributed source of cosmic rays with a similar energy production
rate and energy spectrum (Eq.~\ref{eq:energyrate} and
\ref{eq:energyspectrum} below) would yield agreement with the
observations.

In order to avoid the risk of being misled by  "curve fitting", we
use the same theoretical model that was discussed in
1995~\cite{crflux}. We assume that extra-galactic protons in the
energy range of $10^{19}$~eV to $10^{21}$~eV are produced by
cosmologically-distributed sources at a rate
\begin{equation}
\frac{d\varepsilon}{dt}\approx 3\times 10^{44} {\rm
erg~Mpc^{-3}~yr^{-1}}, \label{eq:energyrate}
\end{equation}
with a power law differential energy spectrum
\begin{equation}
\frac{dN}{dE_p} \propto E_p^{-n}\,,\quad n\approx 2.
\label{eq:energyspectrum}
\end{equation}
We shall refer to this energy spectrum as ``the extra-galactic component" in order to emphasize that the
fit to the data is generic, independent of the type of source that generates the assumed energy and
spectrum. An energy spectrum similar to the assumed energy spectrum, Eq.~(\ref{eq:energyspectrum}), has
been observed for non-relativistic shocks~\cite{spectrum} and for relativistic shocks~\cite{relativistic}
shocks. This power law is produced by Fermi acceleration in collisionless shocks~\cite{spectrum},
although a first principles understanding of the process is not yet available (see, e.g.
Ref.~\cite{arons} for a discussion of alternative shock acceleration processes).

We can use Eq.~(\ref{eq:energyrate}) and Eq.~(\ref{eq:energyspectrum}) to obtain a value for the
cosmological rate, $E_p^2d\dot{N}/dE_p$, at which energy in the form of high energy protons is being
produced. Integrating  $E_pd\dot{N}/dE_p$ between $10^{19}$ eV and $10^{21}$ eV and setting the result
equal to the value given in Eq.~(\ref{eq:energyrate}), we find the proportionality constant in
Eq.~(\ref{eq:energyspectrum}). Thus $ E_p^2d\dot{N}/dE_p ~\approx ~0.7\times10^{44}{\rm
erg\,Mpc{^{-3}}\,yr^{-1}}$.

Energy losses due to pion or pair production are included in the transport calculations in the usual
continuous approximation. Energy loss due to the cosmological redshift is significant for energies
$<5\times10^{19}$~eV, and is also taken into account. The choice of cosmological model is unimportant for
cosmic ray energies above $10^{19}$ eV, which is the region of interest. We assume, for definiteness, a
flat universe with $\Omega_m = 0.3$ and $\Omega_{\Lambda} = 0.7$, and Hubble constant $H_0=65{\,\rm
km/s\,Mpc}$. For consistency with our earlier derivation of the upper bound on neutrino fluxes that
follows from the observed cosmic ray spectrum~\cite{robustupperbound}, we assume that the source density
evolves with redshift $z$ like the luminosity density evolution of QSOs \cite{QSO}, which may be
described as $f(z)=(1+z)^{\alpha}$ with $\alpha\approx3$ \cite{QSOl} at low redshift, $z<1.9$, $f(z)={\rm
Const.}$ for $1.9<z<2.7$, and an exponential decay at $z>2.7$ \cite{QSOh}. This functional form of $f(z)$
is also similar \cite{QSO} to that describing the evolution of star formation rate \cite{SFR}, and also
believed to describe the redshift evolution of GRB rate (see, e.g. \cite{GRBrev} for review). As
mentioned above, the choice of redshift evolution does not affect the spectrum above $10^{19}$ eV.

The cosmic-ray spectrum flattens at $\sim10^{19}$~eV \cite{fly,agasa}. There are indications that the
spectral change is correlated with a change in composition, from heavy to light
nuclei~\cite{fly,composition,HiResMIA}. These characteristics, which are supported by analysis of Fly's
Eye, AGASA and HiRes-MIA data, and for which some evidence existed in previous
experiments~\cite{Watson91}, suggest that the cosmic ray flux is dominated at energies $< 10^{19}$~eV by
a Galactic component of heavy nuclei, and at UHE by an extra-Galactic source of protons. Also, both the
AGASA and Fly's Eye experiments report an enhancement of the cosmic-ray flux near the Galactic disk at
energies $\le10^{18.5}$~eV,  but not at higher energies~\cite{anisotropy}.

We therefore add an observed Galactic component,
\begin{equation}
\frac{dN}{dE} ~\propto~ E^{-3.50}, \label{eq:galacticspectrum}
\end{equation}
to the extra-galactic spectrum component given in
Eq.~(\ref{eq:energyspectrum}). The shape of the energy spectrum of
the Galactic component, Eq.~(\ref{eq:galacticspectrum}), was
derived by the Fly's Eye collaboration~\cite{fly}.

The observed $E^{-2.6}$ spectrum between $1\times 10^{19}$ eV to
$5\times 10^{19}$ eV is, in this model, the combination of two
different source spectra. First, the cosmological distribution of
sources generates an  $E^{-2}$ spectrum (see
Eq.~\ref{eq:energyspectrum}), which energy losses due to
interactions with the CMB steepen to an observed spectrum that is
a bit shallower than $E^{-2.6}$. Second, the Fly's Eye fit to the
Galactic heavy nuclei component  makes a small contribution at
energies $>1\times10^{19}$ eV and is steeper than $E^{-2.6}$. We
will now compare the model spectrum produced by these two sources
with the cosmic ray observations.

\section{Comparison of Model with Cosmic Ray data.}
\label{sec:comparison}

\begin{figure}[!t]
\centerline{(a)\psfig{figure=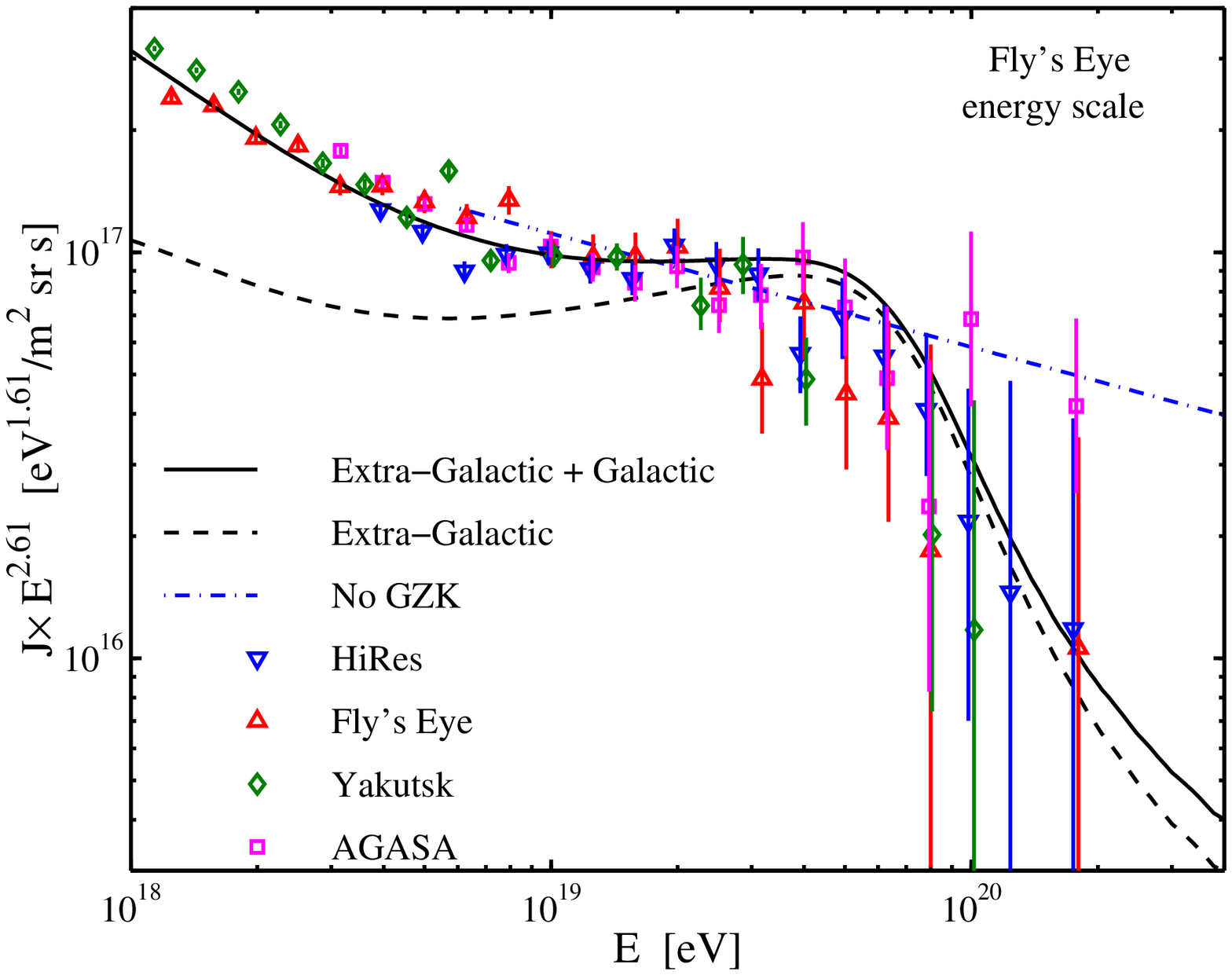,width=3.2in}}
\centerline{(b)\psfig{figure=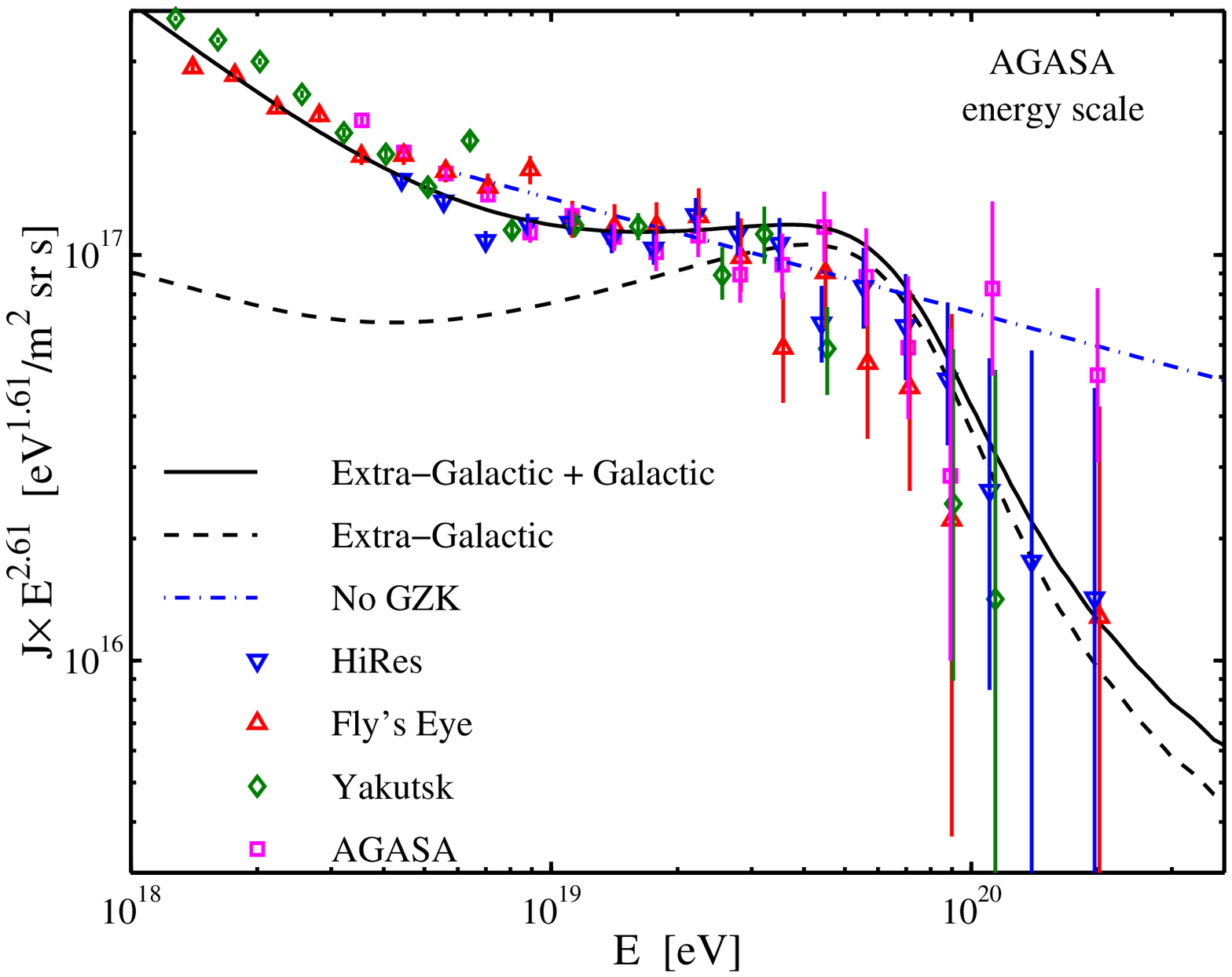,width=3.2in}}
\centerline{(c)\psfig{figure=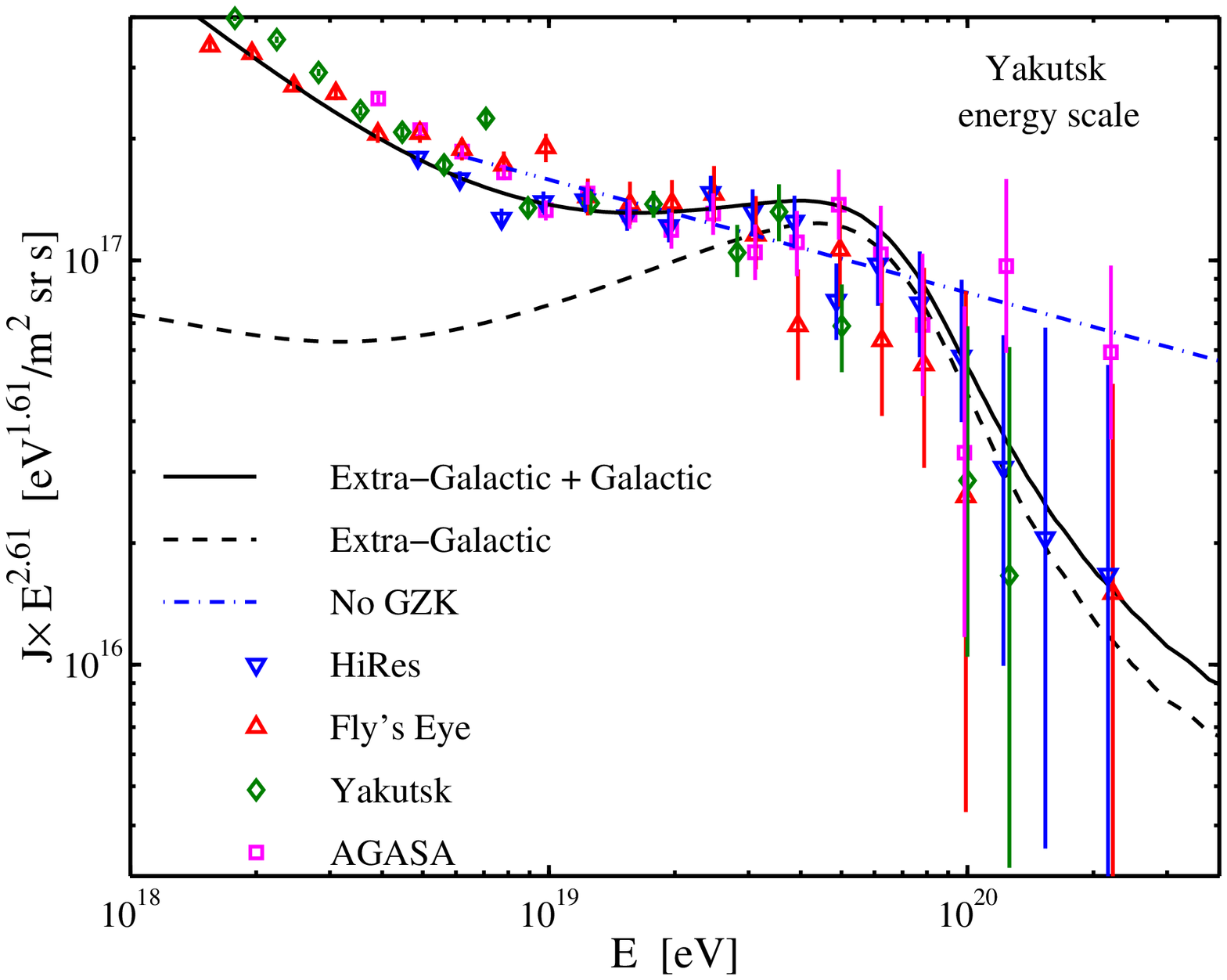,width=3.2in}} \caption{Model versus data. The solid curve shows
the energy spectrum derived from the two-component model discussed in Section~\ref{sec:model}. The dashed
curve shows the extra-Galactic component contribution. The "No GZK" curve is an extrapolation of the
$E^{-2.75}$ energy spectrum derived for the energy range of $6\times 10^{18}$~eV to $4\times 10^{19}$~eV
(\cite{NaganoWatson}; see text) . Three choices of the absolute energy scale are illustrated.
\label{fig:spectrumvsmodel}}
\end{figure}

Figure~\ref{fig:spectrumvsmodel} compares the model prediction with the data from the AGASA~\cite{agasa},
Fly's Eye~\cite{fly}, Hires~\cite{HiRes}, and Yakutsk~\cite{yakutsk} cosmic ray experiments. In order to
demonstrate that our results are insensitive to the choice of absolute energy scale, we present results
for three different choices of the absolute energy scale: adopting the Fly's Eye, the AGASA or the
Yakutsk energy calibration. The three best fit (solid) curves correspond to energy generation rates (see
Eq.~\ref{eq:energyrate}) of $d\varepsilon/dt=\{2.5,3.0,3.5\}\times10^{44}{\rm erg/Mpc^3yr}$ and spectral
indices, (see Eq.~\ref{eq:energyspectrum}) of $n=\{-2.2,-2.1,-2.0\}$ for the \{Fly's Eye, AGASA,
Yakutsk\} energy scales, respectively.

\subsection{Good agreement below $10^{20}$ eV}
\label{subsec:agreementbelow}

The model predictions are in good agreement with the data of all
experiments in the energy range $10^{19}$~eV to $10^{20}$~eV, a
region in which the extra-galactic component is predicted to be
dominant. Since the Fly's Eye representation of the Galactic
component is intended to describe the lower energies, it is not
surprising that the model results are also in good agreement with
the observed spectrum for energies below $10^{19}$~eV.

A "$\chi$-by-eye" comparison of the model to the data shown in Figures~\ref{fig:spectrumvsmodel} appears
to indicate a quantitatively good fit, but we cannot simply compute a formal $\chi^2$ fit due to the
uncertainties in the absolute energy calibration. Instead, we compare the variance of model predictions
from the combined AGASA, Fly's Eye, HiRes and Yakutsk data sets ($s^2_{\rm Model}$) with the variance of
AGASA, Fly's Eye, Haverah Park, and Yakutsk data sets from the HiRes data set ($s^2_{\rm HR}$).  Let
$s^2_{\rm Model}\equiv N^{-1}\sum_{i,j}(n_{ij}-n_{ij,\rm Model})^2/n_{ij,\rm Model}$ where $n_{ij,\rm
Model}$ is the predicted average number of events in the $i$-th energy bin of the $j$-th experiment and
$N$ is the number of bins. Also, let $s^2_{\rm HR}\equiv \tilde{N}^{-1}\sum_{i,j}(n_{ij}-n_{ij,\rm
HR})^2/n_{ij,\rm HR}$, where $n_{ij,\rm HR}$ is the predicted average number of events (for AGASA, Fly's
Eye and Yakutsk) in a model where the HiRes value is the average number of events. We find $s^2_{\rm
HR}=1.06$ with $\tilde{N}=26$ data points, and $s^2_{\rm Model}=1.20$ with $N=36$ data points for all
three choices of the absolute energy scale (panels a, b or c). The different experiments are in agreement
with each other and with the model in the energy range $10^{19}$~eV to $10^{20}$~eV.

\subsection{What is happening above $10^{20}$ eV?}
\label{subsec:above}

\begin{table}[!t]
\centering \caption{Evidence for a GZK suppression. The table compares the number of events expected
above $10^{20}$ eV assuming that there is no GZK suppression with the observed number of events for
different choices of the absolute energy scale. The expected number of events is calculated assuming that
the power law $J\propto E^{-2.75}$ that dominates between $6\times 10^{18}$ eV and $4\times10^{19}$ eV
\cite{NaganoWatson} extends beyond $5\times 10^{19}$ eV. The numbers of events are given for the combined
exposure of the Fly's Eye, HiRes, and Yakutsk experiments. \label{tab:predictedobserved}}
\begin{tabular}{ccc}
\hline\hline \noalign{\smallskip}
Energy Scale&Expected&Observed\\
\noalign{\smallskip} \hline \noalign{\smallskip}
Fly's Eye&34&4\\
AGASA&40&6\\
Yakutsk&46&6\\
\noalign{\smallskip} \hline\hline
\end{tabular}
\end{table}

Above $10^{20}$~eV, Fig.~\ref{fig:spectrumvsmodel} shows that the Fly's Eye, HiRes and Yakutsk
experiments are in agreement with each other and the model. However, the eight AGASA events with energies
greater than $10^{20}$~eV disagree with the prediction of the cosmological model (defined by
Eq.~\ref{eq:energyrate} and Eq.~\ref{eq:energyspectrum}), including the GZK suppression. The Fly's Eye,
Yakutsk and HiRes experiments have a combined exposure three times that of the AGASA experiment.  The
exposures above $10^{20}$~eV are, in units of $10^3{\rm km^2-yr-sr}$: AGASA (1.3), Fly's Eye (0.9),
Yakutsk (0.9), and HiRes (2.2). Together, Fly's Eye, Yakutsk, and Hi-Res observe a total of 6 events
above $10^{20}$ eV (4 events if the Fly's Eye energy scale is chosen).

Assuming no GZK suppression, Table~\ref{tab:predictedobserved} compares the expected number of events
above $10^{20}$ eV with the number of events observed in the combined Fly's Eye, HiRes, and Yakutsk
exposure. The differential energy spectrum observed by the various experiments at the energy range of
$4\times 10^{17}$~eV to $4\times 10^{19}$~eV can be fitted by a broken power-law, where the shallower
component dominating above $\sim6\times 10^{18}$~eV satisfies $J\propto E^{-2.75\pm0.2}$ (see table V and
Eq. 43 in \cite{NaganoWatson}). The expected number of events in the absence of a GZK suppression was
calculated by assuming that the cosmic ray spectrum follows the power law $J\propto E^{-2.75}$ also at
energies $>4\times 10^{19}$~eV. Thus there is a $>5\sigma$ deficit beyond $10^{20}$ eV relative to the
extrapolated lower-energy spectral energy distribution. Adopting the steepest allowed slope, $J\propto
E^{-2.95}$, the expected number of events is \{21,25,30\}, implying a $>3.7\sigma$ deficit beyond
$10^{20}$ eV .

\section{Discussion.}
\label{sec:discussion}

Our most important conclusion is that exotic new physics is not required to account for the observed
events with energies in excess of $10^{20}$ eV, except for the AGASA data.
Table~\ref{tab:predictedobserved} shows that there is already a strong suggestion, $> 5\sigma$
($>3.7\sigma$, depending upon the extrapolated energy spectrum) in the Fly's Eye, HiRes, and Yakutsk
observations that the expected GZK suppression has been observed (see also Fig.~\ref{fig:beforeafter} and
Fig.~\ref{fig:spectrumvsmodel}).

Precision measurements from $10^{18}$~eV to $5\times 10^{19}$~eV are essential for testing models of UHE
cosmic rays, although they are less dramatic than measurements above $10^{20}$~eV. At energies $>
10^{20}$ eV, the predicted number, $N$, of events in conventional models is uncertain due to the unknown
clustering scale, $r_0$, of the sources, $\sigma(N_{\rm predicted})/N_{\rm predicted} = 0.9(r_0/{10~\rm
Mpc})^{0.9}$ ~\cite{clustering}. Paradoxically, we may need to study carefully cosmic rays with energies
below the GZK suppression in order to understand better the origin of the cosmic rays beyond the
suppression.

\section*{Acknowledgments}

We are grateful to  M. Teshima for valuable comments. JNB
acknowledges NSF grant No. 0070928 and the WIS Einstein Center for
hospitality.  EW thanks the IAS for hospitality. EW is the
incumbent of the Beracha foundation career development chair.

\end{document}